# MACROECONOMIC INSTABILITY AND FISCAL DECENTRALIZATION: AN EMPIRICAL ANALYSIS

**Ahmad Zafarullah Abdul Jalil, Mukaramah Harun, Siti Hadijah Che Mat\***

**Abstract:**

The main objective of this paper is to fill a critical gap in the literature by analyzing the effects of decentralization on the macroeconomic stability. A survey of the voluminous literature on decentralization suggests that the question of the links between decentralization and macroeconomic stability has been relatively scantily analyzed. Even though there is still a lot of room for analysis as far as the effects of decentralization on other aspects of the economy are concerned, we believe that it is in this area that a more thorough analyses are mostly called for. Through this paper, we will try to shed more light on the issue notably by looking at other dimension of macroeconomic stability than the ones usually employed in previous studies as well as by examining other factors that might accentuate or diminish the effects of decentralization on macroeconomic stability. Our results found that decentralization appears to lead to a decrease in inflation rate. However, we do not find any correlation between decentralization with the level of fiscal deficit. Our results also show that the impact of decentralization on inflation is conditional on the level of perceived corruption and political institutions.

**Keywords:** decentralization, macroeconomic stability, fiscal federalism, institutional and political environment

**JEL Classification:** E31, H5 H6 H7

## 1. Introduction

The wave of decentralization process has taken into the world since the last two decades. Even though the debates on the benefits of decentralization are yet to be conclusive, as witnessed by the huge literature on this area, it has not stopped various countries from proceeding with the decentralization of their economies. While it cannot be denied that there are countries which have benefited from this process, there are also those whose experiments with decentralization have turned sour.

One area in which studies have been scant is the effects of decentralization on macroeconomic stability. It is also interesting to note that the question of macroeconomic stability itself has been exiguously analyzed in the literature. [1] Macroeconomic instability is usually understood as a situation of economic malaise where the economy does not seem to have settled in a steady position and where, eventually, something needs to be done for putting it back on track (Azam, 2001). To put it differently, macroeconomic instability refers to phenomena that make the domestic macroeconomic environment less predictable and it can take the form of volatility of key macroeconomic variables or of unsustainability in their behaviour. As decentralization implies that more fiscal powers as well as responsibilities will be devolved to subnational governments, the federal government may lose some of its control on the macroeconomic environment of the country. And depending on the behaviour of the subnational governments as well as the level of influence that can be exerted by the federal government on the subnational governments, the macroeconomic conditions of the country will be more or less volatile. In other words, the question of the effect of fiscal decentralization on macroeconomic instability is very much empirical in nature and may vary according to the institutional and political structure of the country in question.

* College of Business, Universiti Utara Malaysia, 06010 Sintok, Kedah, Malaysia (zafar@uum.edu.my)

It is thus the aim of this paper to shed more light on this link between decentralization and macroeconomic instability. The paper starts by reviewing the literature to show that there is yet to be a consensus both at the theoretical and empirical level, among economists as far as the effects of decentralization on macroeconomic instability are concerned. Section 3 presents the data and the methodology employed. The results of our econometric estimations are discussed in Section 4. Finally, Section 5 concludes.

## 2.  The Relationship between Macroeconomic Stability and Decentralization

There are two contrasting beliefs regarding the impact of fiscal decentralization on macroeconomic stability. [2] The first one holds the view that the two are negatively related. With fiscal decentralization, subnational governments are granted more power in determining the level of their expenses as well as their revenues resulting in the multiplicity of political and budget centres. This in turn will make the fiscal management system become more complicated especially the oversight system which is crucial for instilling efficient budget management. In addition to this, subnational governments usually have their own agendas to pursue that may be different from the ones pursued by the federal government thus leading to a coordination problem. Furthermore, there is also the problem of soft budget constraint[3] that underlies the tendency of fiscal profligacy among subnational governments. The increase in autonomy accorded to lower level governments may increase the incentives for opportunistic behaviour among them which may then try to shift their expenditure burdens onto the nation as a whole. Under these circumstances, the coordination of national fiscal and monetary policies as adjustment tools is complicated, posing a challenge to national economic stability (Prud'homme, 1995).

---

1    As it is put by Satyanath & Subramaniam (2004), "*It is surprising that while so much of the recent literature has been devoted to, even obsessed with, explaining the cross-country variation in real variables—for example, in income…there has been much less of a concern with analyzing the cross-country variation in nominal or macroeconomic instability. This is despite the fact that the cross-country variation in nominal or macroeconomic instability is even more astounding than that in income.*" (Satyanath & Subramaniam, 2004: p. 2).

2    It should be noted that no theoretical framework has been developed in order to show the link between the two variables.

On the other hand, under the logic of commitment problem, decentralization is associated with more price and macro stability. In the literature, high inflation is attributed to the inability of policymakers to commit credibly to monetary restraint which in turn is due to the fact that high inflation, regardless of its costs, is their dominant strategy (Barro, Gordon, 1983; Kydland, Presscott, 1977). However, fiscal decentralization through partial devolution of control over spending and monetary policy to lower levels of governments may make it more difficult for policymakers to renege on their commitment for price stability. Furthermore, the competition among lower levels of governments for investments may reduce their incentives to renege on stable monetary policy (Qian, Rolland, 1998).

Another theory suggests that as fiscal decentralization leads to a clear revenue sharing mechanism between state and federal government, there will be less competition for fiscal resources among them. And this will have a positive impact on macroeconomic stability because fiscal competition among different levels of government has been shown to undermine national fiscal policy objectives, particularly by promoting pro-cyclical fiscal policy (Thornton, 2007).

The positive effect of fiscal decentralization on price stability can also take place through the effect that fiscal decentralization has on the independence of central bank. Shah (2005) argued that with fiscal decentralization the central bank will be more independent since a decentralized system would require a more clarified rules and regulations under which a central bank operates as well as its functions and its relationships with different level of governments. Lohmann (1998) argued that Germany's low inflation in the post-war can be partly attributed to the independence of the Bundesbank which was enhanced by the way it was embedded into the country's federal institutions. According to the author, a majority of the bank's council members were appointed by the Lander governments. The Landers were also represented in the Bundesrat which could veto changes to central bank legislation. All those factors serve as checks and balances on the attempts by the central government to inflate the economy in order to gain popularity during elections.

On the empirical grounds, there have been very few studies that analyze the effects of decentralization on macroeconomic stability and almost all of them used the inflation rate as their indicator for macroeconomic instability. King and Ma (2001) found a significant positive correlation between centralization and inflation when they omitted from their sample "high-inflation" countries defined here as those having an average inflation of more than 20 %. They also found that the inclusion of centralization in their regression gives central bank independence the right sign. The paper was later reinvestigated by Neyapti (2003) who took both local accountability, as a fiscal disciplinary device and central bank independence, as a proxy for monetary discipline, into account to assess the relationship between decentralization and inflation. His empirical investigation demonstrates that, controlling for business cycles, openness and government size, revenue decentralization has significant negative effect on inflation only in low inflation countries. These results are consistent with King and Ma's observation of the significant effect of central bank independence. Neyapti observed, however, that decentralization has a significant negative effect on inflation also in higher inflation countries when coupled with both central bank independence and local accountability. More recently, Martinez-Vazquez and McNab (2005) found that decentralization appears to promote price stability. Their results are consistent both in the full and sub-sample of developed, developing and transitional countries. This suggests according to the authors that their results are not dependant on the level of development.

---

3   The soft-budget constraint problem refers to the fact that federal transfers to subnational governments are based on *ex post* financial needs and not, as it should be, on *ex-ante* characteristics of the recipient states. As such, subnational governments are not held to a fixed budget but find their budget constraint softened by the injection of additional credit (or guarantees) whenever they are on the verge of fiscal fiasco. See Rodden (2005) for further explanation of the issue.

Treisman (2000) found that fiscal decentralization have no significant correlation with inflation. The author used three indictors of decentralization, namely, whether the country is classified as federal according to Elzaar (1985), the share of subnational spending of the total government spending and the share of subnational revenue of the total government revenue. The results were confirmed by Rodden and Wibbels (2002). The authors found that although there is a positive correlation between fiscal decentralization and inflation, the relation does not achieve statistical significance. Thornton (2007) found that the impact of revenue decentralization on inflation is not statistically significant. According to Thornton, these results suggest that countries that shift a large share of revenues to sub-national governments are able to pursue better policies at the national level and not a reflection of relatively more responsible fiscal policies at the level of sub-national governments.

Based on the literature review above, it is quite obvious to see that there are still lots to be done as far as the links between decentralization and macroeconomic stability are concerned. Not only that the studies have failed to acknowledge other aspects of macroeconomic stability, they seemed to ignore various other factors that could influence directly or indirectly the effects of decentralization on macroeconomic stability. One type of variables that we believe to be rather important is the ones that capture the existing institutional and political arrangements of the country in question such as the quality of the government, democracy, political stability or the level of corruption. As shown by our review of literature in the preceding section, these variables have been widely studied notably from the angle of the impact that decentralization may have on them. However, there are yet any studies that try to examine the impact that these variables may have on the impact of decentralization on macroeconomic stability. It is not too farfetched to assume for example that the impact of a decentralization process on macroeconomic stability will in a way depend on the position of the country in question in the governance index level. These institutional variables may be introduced into the framework either directly or indirectly through their interaction with other more traditional independent variables. Such interactions may have been widely covered in other studies of decentralization but they are yet to be introduced into the regressions between decentralization and macroeconomic stability.

## 3. Data Description and Econometric Estimation

## 3.1 Data description

Our data covers 62 countries from the period of 1972 to 2001. This data set is structured as a panel with observations for each country consisting of five-year averages. Each country has six observations: 1972-1976, 1977-1981, 1982-1986, 1986-1991, 1992-1996, 1997-2001. The panel is, however, not balanced because some observations are missing for a number of countries. Table 1 below summarizes the descriptive statistics of the variables. Data related to decentralization are obtained from the Government Finance Statistics (GFS) which are collected and published by the International Monetary Fund. More precisely the following indicators will be used as our measure of decentralization

– the percentage of subnational governments expenditure of the total government expenditures
– the percentage of subnational governments revenues of the total government revenues.

The usage of GFS data to measure the extent of decentralization has been widely criticized in the literature.[4] This has led to the use of other type of indicators as well as to the construction of new database by some authors. However compelling the use of these data may be, it will not serve our purpose here as they are only available for certain developed countries. In this study we will use two variables as our measure of macroeconomic instability namely (1) the inflation rate and (2) the level of fiscal deficit.

We have introduced in our estimation the institutional and political setting of a country using the indicators provided by the Center for International Development and Conflict Management under their Polity IV Project.[5] As our measure of corruption, we use the

---

4    See for example Ebel and Yilmaz (2000).

5    The Polity IV data set contains coded annual information on regime and authority characteristics for all independent states in the global state system and covers the years 1800-2004. We will also use the data published by the Freedom House. The latter has over the last 35 years conducted surveys in more than 193 countries and each country was assigned two numerical ratings, one for political rights and one for civil liberties. Each of this rating is based on a 1 to 7 scale. Until 2003, countries whose combined average ratings for Political Rights and for Civil Liberties fell between 1.0 and 2.5 were designated "Free"; between 3.0 and 5.5 "Partly Free," and between 5.5 and 7.0 "Not Free." Beginning with the ratings for 2003, countries whose combined average ratings fall between 3.0 and 5.0 are "Partly Free, and those between 5.5 and 7.0 are "Not Free."

Transparency International historical corruption index.[6] We have also included an indicator of trade openness measured as the percentage of total trade to GDP to control for the impact of openness on price levels as well as the rate of inflation as argued by Romer (1993) and Rogoff (2003).

Table 1
**Descriptive Statistics**

|  | Observations | Mean | Standard Deviation | Min. | Max. |
|---|---|---|---|---|---|
| **Inflation** | 287 | 0.85 | 0.23 | -0.52 | 3.56 |
| **Deficit** | 288 | -3039.67 | 15395.99 | -110922.4 | 81383.6 |
| **Expenditure decentralization** | 261 | 26.286 | 17.0294 | 1.45 | 75.74 |
| **Revenue decentralization** | 252 | 19.04 | 15.30 | 0.62 | 76.40 |
| **GDP** | 287 | 9927.86 | 9234.27 | 137.54 | 41559.24 |
| **M2** | 210 | 61.53 | 212.34 | 1.53 | 1852.05 |
| **Population** | 312 | 6.55e+07 | 1.85e+08 | 164949.6 | 1.25e+09 |
| **Openness** | 287 | 68.16 | 42.48 | 8.68 | 258.47 |
| **Central bank independence** | 131 | 0.46 | 0.19 | 0.17 | 0.89 |
| **Corruption** | 186 | 6.08 | 2.44 | 0.2 | 9.80 |
| **Democracy** | 265 | 5.35 | 7.62 | -52.8 | 10 |
| **Executive constraints** | 265 | 4.27 | 6.64 | -52.4 | 7 |
| **Polity** | 265 | 3.54 | 9.42 | -56.8 | 10 |
| **Political rights** | 303 | 2.67 | 2.02 | -1.8 | 7 |
| **Government size** | 277 | 57353.82 | 123153.3 | 0 | 862603.4 |

## 3.2 Econometric specification

Using the data described above, we estimate the following model

$$MS_{it} = \beta_1 FD_{it} + \beta_2 Pol_{it} + \delta' Z_{it} + u_{it} \qquad (1)$$

where $MS_{it}$ is the measure of macrostability represented here by the inflation rate and the fiscal deficit. Following Neyapti (2003), we use a linear transformation of the rate of inflation that scales it down to the range between zero and one. The formula used for the linear transformation is as follows:

$$Inf = [inflation\ rate/1+inflation\ rate] \qquad (2)$$

---

6    This historical data on the degree to which business transactions involve corruption are reported by the Center of Corruption Research at the University of Groningen jointly with Transparency International. The index is calculated as averages of corruption rankings from Business International, Political Risk Services, World Competitiveness Report, and Political & Economic Risk Consultancy. The index ranges from 0 to 10 with 10 indicating least corruption.

The transformation will allow us to control for the large variance in inflation across countries and over time. $FD_{it}$ is the measure of fiscal decentralization, $Pol_{it}$ denotes a measure of political institutions which will be represented by four variables: the level of political rights, the democracy level, the policy and the constraint on executive. The $Z_{it}$ matrix comprises of several control regressor (M2 as a percentage of GDP, the index of central bank independence, the *per capita* GDP, the total population, the government size, the openness to international trade and the level of corruption). And finally $u_{it}$ is the error term.

We start by testing for the presence of endogeneity problem in our estimation. In the case of our main independent variables, the results show that we fail to reject the exogeneity of the fiscal decentralization with respect to all our dependant variables. We also fail to reject the null hypothesis of exogeneity of several of our control variables namely the GDP *per capita*, the M2 and the openness variable.

We then examine whether a fixed or a random effect model is more appropriate for the estimation of Equation 1.[7] The test for both the rate of inflation and the fiscal deficit are in favour of a random effect specification. Consequently, we used the random effect model in estimating our model.

## 4. Estimation Results

### 4.1 Baseline regressions

The results of our baseline estimations are presented in Table 2 to Table 4. In Table 2, the dependant variable is the inflation rate and fiscal decentralization is measured as the proportion of subnational expenditures to total expenditures. In column A (Table 2), we estimate our regressions without controlling for corruption and political institutions. The results show that there is no significant relationship between inflation and decentralization. Inflation rate appears not to be influenced by decentralization.

In column B (Table 2), we introduce in our regression a variable representing the level of corruption. It is quite striking to see that once we control for corruption, the rate of inflation becomes significantly correlated with the level of expenditure decentralization. The estimated coefficient for expenditure decentralization is found to be significant at the 1% level. The results suggest that in contrary to popular belief of a negative effect of decentralization on macrostability, an increase in expenditure decentralization, all else being equal, would lead to a decrease in the inflation rate. As for the level of corruption, it is also found to be highly correlated with the rate of inflation. An increase in the level of perceived corruption will lead to an increase in the level of inflation.

---

7   The null hypothesis of the Haussman (1978) test is that, assuming that both OLS and GLS are consistent, OLS is inefficient, the alternative being OLS is consistent but GLS is not. In other words, the Haussman statistic tests for the correlation between the individual effects and explanatory variables. Rejection of the null hypothesis thus leads to the rejection of random effects model, in favour of fixed effects (see Hsiao, 1986 or Baltagi, 1995).

These results are suggestive to the importance of taking into account institutional and political settings in assessing the impact of decentralization. In order to further verify this, we include in our regressions several other variables that are supposed to capture the institutional and political context of a country. The results are reported in column C to F (Table 2). However, as we can see from Table 2, none of these four variables appear to have a significant relationship with the rate of inflation. Nevertheless, the inclusion of these political variables does not alter the correlation between decentralization and inflation. The latter remains negatively correlated with decentralization.

As for other control variables, M2 are found to be significantly related to the rate of inflation. The latter is found to be positively related to M2. And once we control for political and institutional variables (column B to F, Table 2), a negative and significant correlation is found between the rate of inflation and the GDP *per capita*. An increase in the level of GDP is thus associated with a decrease in the level of inflation. The results are similar to the one found in other studies (Neyapti 2003; Martinez-Vazquez & McNab, 2005; Thornton, 2007).

In Table 3, we replace our measure of fiscal decentralization with that of the proportion of subnational revenue of the total revenue. Our results are somehow similar to the ones found previously. However, in contrary to expenditure decentralization, revenue decentralization appears to be negatively correlated with the inflation rate even if we do not control for corruption or political institutions. Nevertheless when we introduce corruption and political institutions in our estimations, the statistical significance of the coefficient for decentralization has improved.

Our second measure of macrostability is the deficit level. The results as reported in Table 4 and 5 show that there is no correlation between the deficit level and decentralization. Even after controlling for political variables, there appears to be no correlation between these two variables. Our results are somehow in contrast to the ones by Neyapti (2006) who found decentralization to have a negative impact on deficit. Our results also point to a negative correlation between the government size and the deficit level which is quite surprising as it signifies that a bigger government will lead to less deficit.

## 4.2 The indirect effects of corruption and political institutions

In previous section, we have seen that our measure of macrostability can be influenced by the level of perceived corruption or by political institutions. It is thus interesting to see if in addition to these direct effects, these same variables will also have an impact on the effect of decentralization on macrostability. It seems natural to argue that a positive effect of decentralization on macrostability will somehow be attenuated if the country is plagued with a serious problem of corruption. In contrary, a country which is free from corruption will be able to fully benefit from the effects of decentralization on macrostability. It can also be argued that a more stable political environment may accentuate the impact of decentralization on macrostability and *vice versa.*

In order to test for the assumptions of an indirect effect of corruption and political institutions on macrostability, we introduce in our equation the interaction term between these variables with our measure of decentralization. The results for each indicator of macrostability are reported in Table 6.

In column A (Table 6), the dependant variable is the inflation rate while decentralization is measured by the proportion of subnational governments' expenditure to total government, expenditure. The results of the estimation show that decentralization has a negative impact on the level of inflation. But none of the coefficients estimates of the interaction term are statistically significant which signifies that the impact of expenditure decentralization on inflation is not influenced by corruption and political institutions. It is noteworthy that even after controlling for the level of perceived corruption has a direct positive impact on the level of inflation.

In column B (Table 6), we use revenue decentralization as our measure of decentralization. We note that the impact of decentralization on level of inflation appears to be attenuated by the level of perceived corruption. The parameter estimates for the interaction term between decentralization and corruption is statistically significant. Note that the level of corruption no longer has a direct impact on inflation. The coefficient for corruption is no longer statistically significant. As for political institution, our results show that they do not have direct nor indirect effect on the level of inflation.

In column C and D (Table 6), we use the deficit level as our dependant variable. Similar to the results found previously, the level of deficit is not influenced by decentralization. And the level of corruption as well as political institutions does not have any impact (directly or indirectly) on the level of deficit.

Table 2
**Expenditure Decentralization and Inflation Rate**

| | A | B | C | D | E | F |
|---|---|---|---|---|---|---|
| **Decentrali-zation** | -0.0020 (0.0021) | -0.0057*** (0.0021) | -0.0070*** (0.0022) | -0.0063** (0.0031) | -0.0063** (0.0031) | -0.0064** (0.0031) |
| **Central bank independence** | -0.0352 (0.1356) | -0.1171 (0.1355) | -0.2026 (0.1432) | -0.0621 (0.1929) | -0.0510 (0.1945) | -0.0657 (0.1925) |
| **M2** | 0.0723*** (0.0203) | 0.0802*** (0.0221) | 0.0699*** (0.0225) | 0.1030*** (0.0252) | 0.1039*** (0.0249) | 0.1023*** (0.0254) |
| **GDP** | -0.0089 (0.0270) | -0.1123*** (0.0397) | -0.1089*** (0.0382) | -0.1872*** (0.0554) | -0.1933*** (0.0562) | -0.1838*** (0.0550) |
| **Openness** | -0.0007 (0.0007) | -0.0004 (0.0008) | -0.0001 (0.0007) | -0.0000 (0.0010) | -0.0000 (0.0010) | -0.0000 (0.0010) |
| **Population** | -0.0020 (0.0198) | 0.0197 (0.0206) | 0.0303 (0.0214) | 0.0342 (0.0305) | 0.0352 (0.0311) | 0.0333 (0.0303) |
| **Corruption** | | -0.0777*** (0.0233) | -0.0675*** (0.0231) | -0.1312*** (0.0317) | -0.135*** (0.0321) | -0.1291*** (0.0318) |
| **Political rights** | | | -.035347 (0.0261) | | | |
| **Democracy** | | | | 0.0016 (0.0029) | | |
| **Polity** | | | | | 0.0014 (0.0025) | |
| **Executive contraints** | | | | | | 0.0016 (0.0032) |
| **Constant** | 0.8448* (0.4979) | 1.0559** (0.5099) | 1.0690** (0.4923) | 1.0359 (0.7398) | 1.0460 (0.7479) | 1.0413 (0.7334) |
| **R2** | 0.3801 | 0.5046 | 0.5942 | 0.3278 | 0.3269 | 0.3291 |
| **No. of observations** | 294 | 294 | 294 | 294 | 294 | 294 |
| **No. of countries** | 49 | 49 | 49 | 49 | 49 | 49 |

Notes: standard error in parentheses; significant at 10% level*, significant at 5% level**, significant at 1% level**

Table 3
**Revenue Decentralization and Inflation Rate**

| | A | B | C | D | E | F |
|---|---|---|---|---|---|---|
| **Decentralization (revenue)** | -0.0045* (0.0026) | -0.0069*** (0.0026) | -0.0081*** (0.0025) | -0.0088** (0.0039) | -0.0089** (0.0039) | -0.0088** (0.0038) |
| **Central bank independence** | -0.0516 (0.1358) | -0.0749 (0.1380) | -0.1491 (0.1401) | -0.0741 (0.1926) | -0.0673 (0.1937) | -0.076 (0.1925) |
| **M2** | 0.0681*** (0.0201) | 0.0892*** (0.0219) | 0.0777*** (0.0222) | 0.1058*** (0.0242) | 0.1061*** (0.0241) | 0.1054*** (0.0245) |
| **GDP** | 0.0059 (0.0279) | -0.0924** (0.0426) | -0.0772* (0.0396) | -0.1631*** (0.0578) | -0.1688*** (0.0586) | -0.1594*** (0.0574) |
| **Openness** | -0.0008 (0.0007) | -0.0002 (0.0008) | 0.0001 (0.0008) | 0.0000 (0.0010) | 0.0000 (0.0010) | 0.0000 (0.0010) |
| **Population** | 0.0058 (0.0198) | 0.0262 (0.0224) | 0.0342 (0.0222) | 0.0452 (0.0317) | 0.0464 (0.0321) | 0.0438 (0.0314) |
| **Corruption** | | -0.0725*** (0.0228) | -0.0551** (0.0228) | -0.1272*** (0.0301) | -0.1312*** (0.0305) | -0.1246*** (0.0300) |
| **Political rights** | | | -0.0314 (0.0259) | | | |
| **Democracy** | | | | 0.0019 (0.0028) | | |
| **Polity** | | | | | 0.0016 (0.0024) | |
| **Executive contraints** | | | | | | 0.0019 (0.0031) |
| **Constant** | 0.6502 (0.5093) | 0.7234 (0.5656) | 0.6994 (0.5269) | 0.6565 (0.7854) | 0.6673 (0.7909) | 0.6670 (0.7793) |
| **R2 between** | 0.4031 | 0.4945 | 0.5833 | 0.3374 | 0.3404 | 0.3364 |
| **No. of observations** | 294 | 294 | 294 | 294 | 294 | 294 |
| **No. of countries** | 49 | 49 | 49 | 49 | 49 | 49 |

Notes: standard error in parentheses; significant at 10% level*, significant at 5% level**, significant at 1% level***.

Table 4
**Expenditure Decentralization and Deficit**

| | A | B | C | D | E | F |
|---|---|---|---|---|---|---|
| **Decentrali-zation** | -249.5539 (210.3611) | -45.9466 (216.7521) | 7.6788 (227.7219) | -122.5515 (220.7684) | -131.4853 (220.4912) | -114.6224 (219.9376) |
| **Government size** | -0.0670*** (0.0130) | -0.0622*** (0.0133) | -0.0610*** (0.0135) | -0.0413** (0.0202) | -0.0422** (0.0202) | -0.0407** (0.0202) |
| **GDP** | 22801.16*** (6636.693) | 19730.12*** (7208.145) | 19423.88*** (7232.238) | 14015.55* (7921.989) | 13835.79* (7906.534) | 14301.21* (7901.63) |
| **Openness** | -97.6126 (77.7238) | -47.5449 (76.3210) | -47.6396 (76.464) | 9.3261 (89.0054) | 13.15467 (88.8271) | 4.6377 (88.7444) |
| **Population** | -11894.04 (9165.708) | -20082.17 (14006.72) | -18892.26 (14115.13) | -20972.66 (15239.68) | -20516.86 (15232.87) | -21202.37 (15190.46) |
| **Corruption** | | 543.5021 (1298.573) | 780.0778 (1335.694) | 42.4183 (1456.478) | -67.1448 (1456.135) | 134.6995 (1447.768) |
| **Political rights** | | | 885.5985 (1132.154) | | | |
| **Democracy** | | | | -8.4391 (142.2871) | | |
| **Polity** | | | | | 55.8371 (115.6731) | |
| **Executive contraints** | | | | | | -111.757 (164.1635) |
| **Constant** | 9679.554 (135345.6) | 161516.9 205175.4 | 139397.8 (207495.2) | 229442.4 (239256.3) | 223910 (239101.2) | 230605.8 (238518.7) |
| **R2 between** | 0.1553 | 0.2435 | 0.2480 | 0.1240 | 0.1266 | 0.1291 |
| **No. of observations** | 294 | 294 | 294 | 294 | 294 | 294 |
| **No. of countries** | 49 | 49 | 49 | 49 | 49 | 49 |

Table 5
**Revenue Decentralization and Fiscal Deficit**

| | A | B | C | D | E | F |
|---|---|---|---|---|---|---|
| **Decentrali-zation** | -192.3795 (137.7678) | -471.4344 (389.9944) | -468.6184 (390.6733) | -418.7051 (392.284) | -439.5725 (391.2985) | -392.293 (391.6007) |
| **Government size** | -0.0552*** (0.0091) | -0.0638*** (0.0132) | -0.0623*** (0.0134) | -0.0472** (0.0204) | -0.0485** (0.0204) | -0.0462** (0.0203) |
| **GDP** | 13257.67*** (4031.76) | 19930.49*** (7113.249) | 19434.69*** (7151.949) | 15111.64* (7864.256) | 15005.43* (7846.61) | 15318.68* (7848.482) |
| **Openness** | -29.6787 (46.167) | -47.1042 (75.7441) | -46.4211 (75.8776) | 0.8634 (88.4786) | 2.2704 (88.2689) | -4.8144 (88.2783) |
| **Population** | -10747.18* (5496.96) | -17430.28 (13818.65) | -15516.25 (14045.13) | -19394.46 (15174.63) | -18879.57 (15161.37) | -19726.99 (15135.67) |
| **Corruption** | | 749.133 (1296.869) | 959.3556 (1325.096) | 364.5793 (1455.059) | 273.8363 (1453.808) | 437.1089 (1448.281) |
| **Political rights** | | | 862.5821 (1072.218) | | | |
| **Democracy** | | | | 1.7443 (141.9599) | | |
| **Polity** | | | | | 63.4994 (115.2619) | |
| **Executive contraints** | | | | | | -98.7543 (164.1183) |
| **Constant** | 64781.99 (80763.88) | 122287 (203851) | 91371.24 (207782.4) | 197243.9 (240619.3) | 190185.6 (240388.5) | 200597.2 (240010.2) |
| **R2** | 0.2113 | 0.2537 | 0.2583 | 0.1331 | 0.1364 | 0.1371 |
| **No. of observations** | 294 | 294 | 294 | 294 | 294 | 294 |
| **No. of countries** | 49 | 49 | 49 | 49 | 49 | 49 |

Notes: standard error in parentheses; significant at 10% level*, significant at 5% level**, significant at 1% level***

Table 6
**Indirect Effects of Corruption and Political Institutions**

|  | Inflation | | Deficit | |
|---|---|---|---|---|
|  | **A** | **B** | **E** | **F** |
| **Decentralization (expenditure)** | -0.0143** (0.0067) |  | -44.0657 (279.3191) |  |
| **Decentralization (revenue)** |  | -0.0243*** (0.0087) |  | -164.9142 (255.2185) |
| **Central bank independence** | -0.0352 (0.1980) | -0.1276 (0.1946) |  |  |
| **M2** | 0.1020*** (0.0250) | 0.0983*** (0.0237) |  |  |
| **GDP** | -0.2104*** (0.0578) | -0.1629*** (0.0581) | 1148.653 (1815.006) | 546.6188 (1156.83) |
|  |  |  | -0.0713*** (0.0091) | -0.0303*** (0.0084) |
| **Openness** | -0.0001 (0.0010) | -0.0002 (0.0010) | -17.4974 (41.2849) | -2.6873 (29.2362) |
| **Population** | 0.0324 (0.0320) | 0.0372 (0.0323) | -1742.2 (1191.689) | -1225.501 (779.5313) |
| **Corruption** | -0.0889* (0.0465) | -0.0581 (0.0475) | -127.8501 (1286.163) | -603.8841 (791.484) |
| **Democracy** | 0.0019 (0.0068) |  | -1.5947 (369.4532) | 347.5146 (241.9974) |
| **Corruption*Dec** | -0.0014 (0.0009) | -0.0025** (0.0012) | 17.0136 (39.0174) | 34.8798 (31.8542) |
| **Demo*Dec** | -0.0000 (0.0001) | -0.0000 (0.0001) | 3.1053 (9.1577) | -5.6113 (7.7528) |
| **Constant** | 1.4690* (0.8220) | 1.2440 (0.8373) | 17321.16 (26545.97) | 15623.28 (17641.03) |
| **R2 between** | 0.3490 | 0.3948 | 0.8045 | 0.4550 |
| **No. of observations** | 294 | 294 | 294 | 294 |
| **No. of countries** | 49 | 49 | 49 | 49 |

Notes: standard error in parentheses; significant at 10% level*, significant at 5% level**, significant at 1% level***.

## 5. Conclusion

The main objective of this paper is to fill a critical gap in the literature on the relationship between decentralization and macroeconomic stability. The empirical results provided in this study despite data inadequacies and methodological shortcoming, point to the fact that there is a negative relationship between certain variable of macrostability and decentralization. In our baseline estimations, we found that decentralization appears to lead to a decrease in inflation rate. However, we do not find any correlation between decentralization with the level of fiscal deficit. The results suggest that decentralization is not disastrous for macroeconomic stability. In contrast, it has a negative impact on inflation and does not deteriorate nor ameliorate the fiscal balance of a country. Our results seem to run counter to a rapidly growing popular belief that decentralization is disastrous to macroeconomic stability. Our results also show that the impact of decentralization on inflation is conditional on the level of perceived corruption and political institutions. Together, these results indicate that the benefit from fiscal decentralization can only be realized if a country possesses the right kind of political or institutional environment.

Our findings are in line with the views on the positive relationship between fiscal decentralization and macroeconomic stability discussed in the previous section. Indeed, all these views implicitly suggest that for fiscal decentralization to have a positive impact on macroeconomic stability, it would require the existence of a good political and institutional setting. For example, the argument made by Thorton (2000) that fiscal decentralization leads to a clear sharing mechanism of fiscal revenues between subnational governments and federal government and therefore less competition for fiscal resources between them may not be totally valid if the process took place in a country which is plagued with serious corruption problem or is run in an undemocratic manner. These results are also suggestive of the fact that the negative impacts of fiscal decentralization on macroeconomic stability as suggested in the literature review may not materialize if the country in question possess a sound institutional and political environment.

As for fiscal deficit, our results seem to suggest that even though state governments may be subject to a soft budget constraint, the extent towards which this may be affecting the nation's total fiscal deficit is still small. In other words, the size of the subnational governments in proportion to the total size of the government is still small and the level of decentralization is still relatively minor therefore, the impact on the national fiscal balance may not be substantial. Further studies should thus look into the issue if there is any level beyond which fiscal decentralization starts to have an impact on fiscal deficit.